\documentstyle[12pt,psfig]{article}

\def\bel#1{\begin{equation}\label{#1}}
\def\ee{\end{equation}}
\def\bea{\begin{eqnarray}}
\def\eea{\end{eqnarray}}

\def\re#1{Eq.~(\ref{#1})}
\def\ds{\displaystyle}
\def\pari{{\scriptscriptstyle\parallel}}
\def\bm#1{\mbox{\boldmath$#1$\unboldmath}}

\def\loo{\,\raisebox{-.5ex}{$\stackrel{<}{\scriptstyle\sim}$}\,}

\def\Gt{\Gamma_{\omega_{\scriptstyle *}}}
\def\Gti{\Gamma (\omega_{\scriptstyle *}\rightarrow i)}

\def\putf#1#2{\begin{minipage}[t]{56mm}
\hspace{-5mm}\psfig{width=50mm,angle=-90,file=#1.ps}\\[-15mm]
#2\end{minipage}}

\def\Journal#1#2#3#4{{#1} {\bf #2},~#3~(#4)}
\def\NPA{{\em Nucl. Phys.} A}
\def\PRC{{\em Phys. Rev.} C}
\def\ZPC{{\em Z.~Phys.} C}

\begin{document}
\baselineskip 24pt
\vspace*{1cm}
\begin{center}
{\Large\bf Particle production
by time--dependent meson fields\\ in relativistic
heavy--ion collisions}\\
{\bf I.N.~Mishustin$^{a,b}$, L.M.~Satarov$^a$, H.~St\"ocker$^c$ and
W.~Greiner$^c$}
\end{center}
\baselineskip 16pt
\begin{tabbing}
\hspace*{4em}\=${}^a$\,\={\it The Kurchatov~Institute,
123182~Moscow,~\mbox{Russia}}\\
\>${}^b$\>{\it The Niels~Bohr~Institute,
DK--2100~Copenhagen {\O},~\mbox{Denmark}}\\
\>${}^c$\>{\it Institut~f\"{u}r~Theoretische~Physik,
J.W.~Goethe~Universit\"{a}t,}\\
\>\>{\it D--60054~Frankfurt~am~Main,~\mbox{Germany}}
\end{tabbing}
\baselineskip 24pt

\begin{abstract}
According to the Walecka mean--field theory of nuclear interaction
the collective mutual deceleration of the colliding nuclei gives rise to the
bremsstrahlung of real and virtual $\omega$--mesons. It is shown that decays
of these mesons may give a noticeable contribution to the observed yields of
the baryon--antibaryon pairs, dileptons and pions. Excitation functions and
rapidity distributions of particles produced by this mechanism are calculated
under some simplifying assumptions about the space--time variation of meson
fields in nuclear collisions. The calculated multiplicities of coherently
produced particles grow fast with the bombarding energy, reaching a
saturation above the RHIC bombarding energy. In the case of central Au+Au
collisions the bremsstrahlung mechanism becomes comparable with particle
production in incoherent hadron--hadron collisions above the AGS energies.
The rapidity spectra of antibaryons and pions exhibit a characteristic
two--hump structure which is a consequence of incomplete projectile--target
stopping at the initial stage of the reaction.  The predicted distribution of
e$^+$e$^-$ pairs has a strong peak at invariant masses
\mbox{$M_{{\rm e}^+{\rm e}^-}<0.5$}~GeV.
\end{abstract}

\section{Introduction}

As follows from the relativistic mean--field model~\cite{Wal85} strong
time--dependent meson fields are generated in the course of relativistic
heavy--ion collisions. Within the framework of this model several new
collective phenomena were predicted: the filamentation instability of
interpenetrating nuclei~\cite{Iva89} and the spontaneous creation of the
baryon--antibaryon ($B\overline{B}$) pairs in a superdense baryon--rich
matter~\cite{Mis90}. Using the approach developed in papers on the
pion~\cite{Vas80} and photon~\cite{Lip88} bremsstrahlung we suggested
recently~\cite{Mis95} a new mechanism of the $B\overline{B}$ pair production
by the collective bremsstrahlung of meson fields in relativistic heavy--ion
collisions. These pairs may be produced at sufficiently high bombarding
energies when characteristic Fourier frequencies of meson fields exceed the
energy gap between the positive and negative energy levels of baryons.

In the lowest order approximation the production of the $B\overline{B}$ pair
may be considered as a two--step process
$A_pA_t\rightarrow\omega_{\scriptstyle *}\rightarrow B\overline{B}$.  Here
$A_p (A_t)$ denotes the projectile (target) nucleus and $\omega_*$ indicates
the off--mass--shell vector meson \footnote{As discussed in
Ref.~\cite{Mis95}, at relativistic bombarding energies bremsstrahlung of the
scalar meson field is small as compared to the vector meson field. Due to
this reason we disregard here the contribution of the scalar meson
bremsstrahlung.}. The first step in the above reaction is the virtual
bremsstrahlung producing virtual mesons with masses $M>2m_B$ ($m_B$ is the
baryon mass).  The second step is the conversion of a vector meson into
$B\overline{B}$ pairs. Such a process is suppressed for a ''real'' $\omega$
meson which has the mass $m_\omega\simeq 0.783$ GeV and the relatively small
width $\Gamma_\omega\simeq 8.4$ MeV.  It is clear that analogous
bremsstrahlung mechanism may produce also pions (by decays of quasireal
mesons $\omega\rightarrow\pi^+\pi^0\pi^-$) and low mass dileptons
($M_{l^+l^-}\loo m_\omega$).

In this work we study the bremsstrahlung of vector meson fields originated
from the collective deceleration of the projectile and target nuclei at
the initial stage of a heavy--ion collision. The various channels of the
bremsstrahlung conversion, including the production of the $N\overline{N}$
pairs, pions and dileptons are considered with emphasize to their observable
signals.

\section{Particle production by bremsstrahlung of nuclear meson
fields}

By the analogy to the Walecka model we introduce the vector meson field
$\omega^\mu(x)$ coupled to the 4--current $J^\mu(x)$ of nucleons
participating in a heavy--ion collision at a given impact parameter.
The equation of motion defining the space--time behavior of $\omega^\mu(x)$
may be written as ($c=\hbar=1$)
\bel{eqm}
(\partial^\nu\partial_\nu+m_\omega^2)\,\omega^\mu(x)=g_V J^\mu(x)\,,
\ee
where $g_V$ is the $\omega N$ coupling constant. In the mean--field
approximation the quantum fluctuation of $J^\mu$ are disregarded and
the vector meson field is purely classical.~From~\re{eqm} one can see
that excitation of propagating waves in a vacuum (bremsstrahlung)
is possible if the Fourier transformed baryonic current
\bel{ftbc}
J^\mu(p)=\int{\rm d}^4x J^\mu(x) e^{ipx}
\ee
is nonzero in the time--like region $p^2=m_\omega^2$\,.

In the following we study the bremsstrahlung process in the lowest order
approximation neglecting the back reaction and reabsorption of the emitted
vector mesons, i.e. treating $J^\mu$ as an external current. From~\re{eqm}
one can calculate the energy flux of the vector field at a large distance
from the collision region~\cite{Vas80}. This leads to the following
formulae for the momentum distribution of real $\omega$--mesons
emitted in a heavy--ion collision~\cite{Iva89}
\bel{rvms}
E_\omega\frac{\ds{\rm d}^3 N_\omega}
{\ds{\rm d}^3p\hfill}=S(E_\omega,\bm{p})\,,
\ee
where $E_\omega=\sqrt{m_\omega^2+\bm{p}^{\,2}}$ and
\bel{sfu}
S(p)=\frac{g_V^2}{16\pi^3}|J_\mu^*(p)J^\mu(p)|
\ee
is a source function. In our model the latter is fully determined
by the collective motion of the projectile and target nucleons.

To take into account the off--mass--shell effects we characterize
virtual $\omega$ mesons by their mass $M$ and total
width $\Gamma_{\omega_{\scriptstyle *}}$. The spectral function describing
the deviation from the on--mass--shell may be written as
\bel{spf}
\rho(M)=\frac{2}{\pi}\frac{\ds M\Gt}
{\ds (M^2-m_\omega^2)^2+m_\omega^2\Gt^2}\,.
\ee
To calculate the distribution of virtual mesons in their 4--momentum
$p$ we use the formulae~\cite{Kno95}
\bel{dvom}
\frac{\ds{\rm d}^4 N_{\omega_{\scriptstyle *}}}
{\ds{\rm d}^4p\hfill}= \rho(M) S(p)\,,
\ee
where $M\equiv\sqrt{p^2}$\,. In the limit
$\Gamma_{\omega_{\scriptstyle *}}\rightarrow 0$ one can replace $\rho(M)$ by
$2\delta(M^2-m_\omega^2)$\,. In this case~\re{dvom} becomes equivalent to the
formulae~(\ref{rvms}) for the spectrum of the on--mass--shell vector mesons.

Below we consider the most important channels of the virtual $\omega$ decay:
$i=3\pi, N\overline{N}$, ${\ds e}^+{\ds e}^-$, $\mu^+\mu^-$\,. The total width
$\Gt$ can be decomposed into the sum over the partial decay widths $\Gti$\,:
\bel{pdw}
\Gt=\sum_i{\Gti}\,.
\ee
The distribution over the total 4--momentum of particles in a given
decay channel may be written as
\bel{dtmi}
\frac{\ds{\rm d}^4 N_{\omega_{\scriptstyle*}\rightarrow i}}
{\ds{\rm d}^4p\hfill}=B(\omega_{\scriptstyle *}\rightarrow i)
\frac{\ds{\rm d}^4 N_{\omega_{\scriptstyle *}}}{\ds{\rm d}^4p\hfill}\,,
\ee
where $B(\omega_{\scriptstyle *}\rightarrow i)\equiv\Gti/\Gt$ is the
branching ratio of the $i$--th decay channel. The latter is a function
of the invariant mass of the decay particles~$M$.

To calculate the 4--vectors $J^\mu(p)$ defining the source function $S(p)$ we
assume the simple picture of a high--energy heavy--ion collision suggested in
Ref.~\cite{Mis95} Below we consider collisions of identical nuclei
($A_p=A_t=A$) at zero impact parameter. In the equal velocity frame the
projectile and target nuclei initially move to each other with the velocity
$v_0=(1-4m_N^2/s)^{1/2}$, where $\sqrt{s}$ is the c.m.  bombarding energy per
nucleon. In the ''frozen density'' approximation~\cite{Mis95} the internal
compression and transverse motion of nuclear matter are disregarded at the
early (interpenetration) stage of the reaction. Within this approximation the
colliding nuclei move as a whole along the beam axis with instantaneous
velocities $\dot{\xi}_p=-\dot{\xi}_t\equiv\xi(t)$\,. The projectile velocity
$\dot{\xi}(t)$ is a decreasing function of time, chosen in the
form~\cite{Vas80}
\bel{prtr}
\dot{\xi}(t)=v_f+\frac{\ds v_0-v_f}{\ds 1+{\rm e}^{t/\tau}}\,,
\ee
where $\tau$ is the effective deceleration time and $v_f$ is the
final velocity of nuclei (at $t\rightarrow +\infty$)\,.

In our approximation the Fourier transforms $J^\mu(p)$ are totally
determined by the projectile trajectory $\xi(t)$~\cite{Mis95}:
\bel{ftre}
J^0(p)=\frac{p_\pari}{p_0}J^3(p)=
2A\int\limits_{-\infty}^\infty{\rm d}t
e^{\ds ip_0t}\cos{[p_\pari \xi(t)]}\,F\left(\sqrt{\bm{p}_T^2+
p_\pari^2\cdot[1-\dot{\xi}^2(t)]}\right)~,
\ee
where $p_\pari$ and $\bm{p}_T$ are, respectively, the longitudinal
and transverse components of the three--momentum $\bm{p}$,  $F(q)$ is
the density form factor of the initial nuclei
\bel{dff}
F(q)\equiv\frac{1}{A}\int{\rm d}^3r\rho(r) e^{\ds-i\bm{q}\,\bm{r}}\,.
\ee
The time integrals in \re{ftre} were calculated numerically assuming
the Woods--Saxon distribution of the nuclear density $\rho(r)$\,. According
to~Eqs. (\ref{sfu}), (\ref{ftre}) the source function $S(p)$ vanishes at
$p_\pari=0$\,.  As a result, at high bombarding energies single particle
distributions have a dip in a central rapidity region (see Figs.~2--3 and
Ref.~\cite{Mis95}). The two--hump structure of the rapidity spectra is a
consequence of the incomplete mutual stopping of nuclei at the initial stage
of a heavy--ion collision. On the other hand, the conventional mechanism of
particle production in incoherent hadron--hadron collisions results in
rapidity distributions of pions and antiprotons with a single central maximum
even at high bombarding energies~\cite{Scho93,Amel95}.

In this work we use the same choice of the coupling constant $g_V$ and
the stopping parameters $\tau, v_f$ as in Ref.~\cite{Mis95} In particular,
it is assumed that $\tau$ equals one half of the nuclear passage time
\bel{tauc}
\tau=R/\sinh{y_0}\,,
\ee
where $R$ is the geometrical radius of initial nuclei. Instead of
$v_f$ we introduce the c.m. rapidity loss $\delta y$:
\bel{vfin}
v_f=\tanh{(y_0-\delta y)}\,.
\ee
In the case of a central Au+Au collision we assume the energy--independent
value~\cite{Scho93} $\delta y=2.4$ for $\sqrt{s}>10$ GeV and full
stopping ($\delta y=y_0$) for lower bombarding energies.

The partial width of the $3\pi$ decay channel is calculated assuming
that $\Gamma (\omega_{\scriptstyle *}\rightarrow 3\pi)$ is proportional
to the three--body phase space volume~\cite{Iva89}. The normalization
constant is determined from the condition that
$B(\omega_{\scriptstyle *}\rightarrow 3\pi)$ equals the observable
value $B(\omega\rightarrow 3\pi)=0.89$ at $M=m_\omega$. Calculation
of the $\omega_{\scriptstyle *}\rightarrow N\overline{N}$ matrix element
in the lowest order approximation in $g_V$ gives the result
\bel{wnnb}
\Gamma (\omega_{\scriptstyle *}\rightarrow N\overline{N})=
\frac{\ds g_V^2}{\ds 6\pi}\sqrt{M^2-4m^2_N}\left(1+\frac{\ds 2m^2_N}
{\ds M^2}\right)\Theta(M-2m_N)\,,
\ee
where $\Theta(x)=\frac{1}{2}(1+{\rm sign}\,x)$\,. After substituting
(\ref{wnnb}) into \re{dvom} and omitting the second term in the denominator
of $\rho(M)$ one arrives at the distribution over the pair 4--momentum
obtained earlier in Ref.~\cite{Mis95}

The dilepton production is studied
by calculating\hspace{1.5ex}the matrix\hspace{1.5ex}elements of the process
\mbox{$\omega_{\scriptstyle *}\rightarrow\gamma_{\scriptstyle *}$}
$\rightarrow l^+l^-$
where $\gamma_{\scriptstyle *}$ is a virtual photon. This calculation
leads to the result ($l={\rm e}, \mu$)~\cite{Koch93}
\bel{wdl}
\frac{\ds \Gamma (\omega_{\scriptstyle *}\rightarrow l^+l^-)}
{\ds \Gamma (\omega\rightarrow l^+l^-)}=\left(\frac{\ds m_\omega}
{\ds M}\right)^6\frac{\ds M^2+2m_l^2}{\ds m_\omega^2+2m^2_l}
\sqrt{\frac{\ds M^2-4m_l^2}{\ds m_\omega^2-4m_l^2}}\,\Theta(M-2m_l)\,,
\ee
where $m_l$ is the lepton mass. 

\section{Results}

Below we present the results of numerical calculations obtained within the
model described in the preceding section. Some of the model predictions
concerning the $B\overline{B}$ production have been already published in
Ref.~\cite{Mis95}\\
\hspace*{1cm}
\putf{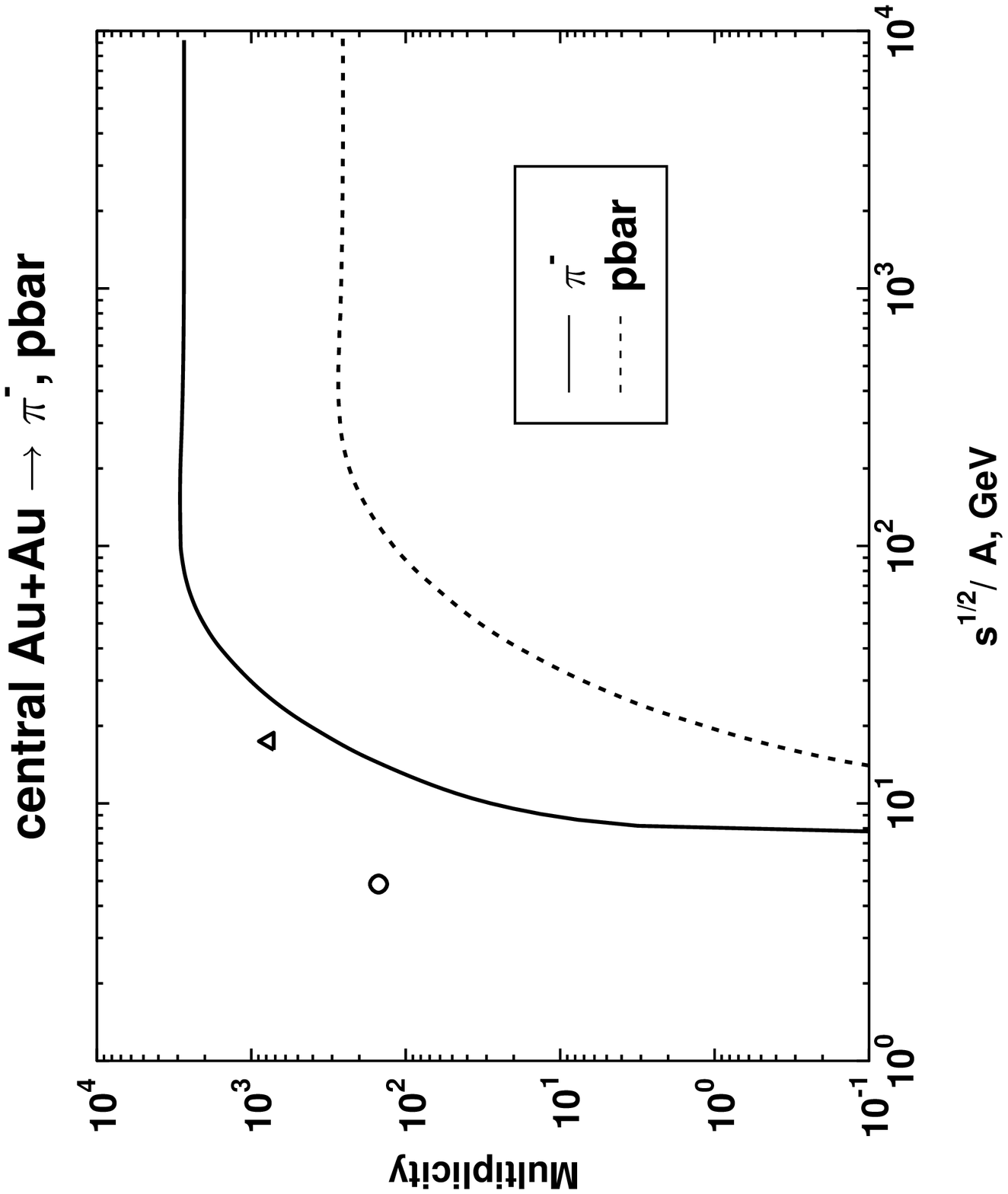}
{Figure~1: Excitation functions of $\pi^-$ mesons (solid line) and
anti\-pro\-tons (dashed line) produced by brems\-strahlung in central
Au+Au collision. Circle and triangle are experimental data on $\pi^-$
multiplicity (see text).}
\hspace{2cm}
\putf{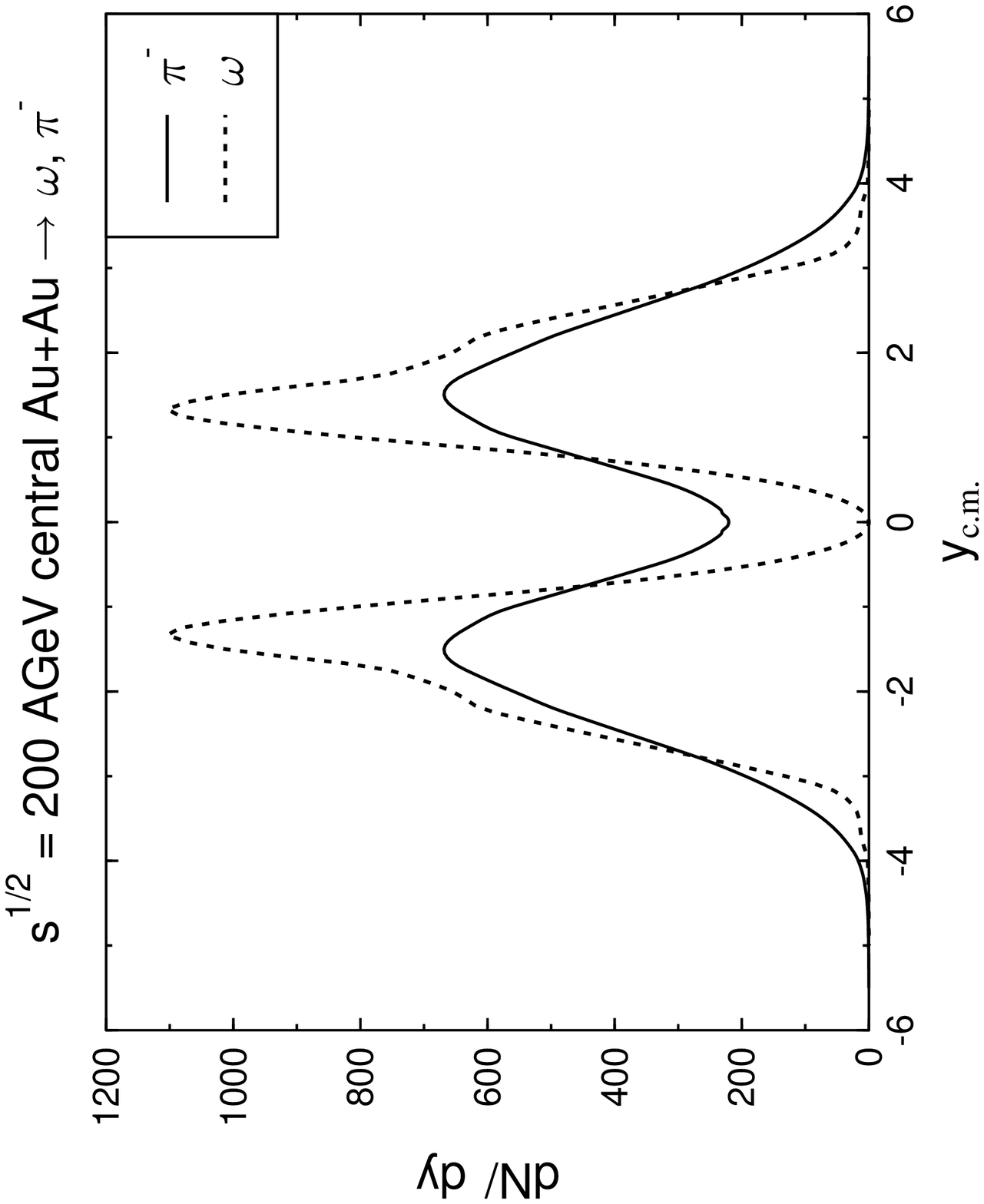}
{Figure~2: Rapidity distributions
of $\omega$ (solid line) and $\pi^-$ (dashed line) mesons produced by
brems\-strahlung in central Au+Au collision at RHIC bombarding
energy.}\\[1cm]
Fig.~1 shows the $\pi^-$ and $\overline{p}$
multiplicities as functions of the bombarding energy in the case of central
Au+Au collisions. For comparison, we show experimental data on the $\pi^-$
multiplicities in the 11.6 AGeV/c Au+Au (circle)~\cite{Gon93} and 160 AGeV
Pb+Pb (triangle)~\cite{Mar95} central collisions. 
One can see that the
multiplicity of pions produced by bremsstrahlung exhibits a rapid growth
between the AGS ($\sqrt{s}\simeq 5$~AGeV) and the SPS ($\sqrt{s}\simeq 20$
AGeV) energies and saturates above the RHIC ($\sqrt{s}\simeq 200$ AGeV)
energy region. It is interesting that the bremsstrahlung component of pion
yield becomes comparable with pion production in incoherent hadron--hadron
collisions~\cite{Amel95} already at the SPS bombarding energies. Note,
however, that actual pion and antiproton yields, especially for heavy
combinations of nuclei, may be reduced due to the absorption and annihilation
neglected in the present model.

The results on the $\pi^-$ rapidity spectra are represented in Figs.~2--3.
The spectra are calculated in the limit of the on--mass--shell $\omega$
mesons, i.e. assuming $\Gt=0$. Here we use the kinematic formulae
connecting the pion spectrum and the ''primordial'' distribution of vector
mesons, \re{rvms}, suggested in Ref.~\cite{Iva89} Similarly to the case of
antiprotons~\cite{Mis95}, the pion rapidity spectrum has a pronounced dip at
$y_{\,\rm c.m.}\simeq 0$. The two--hump structure of the $\pi$ and
$\overline{p}$ spectra may serve as a signature of the bremsstrahlung
mechanism. According to Fig.~3 this structure can be seen only at high enough
bombarding energies.


\hspace{1cm}\putf{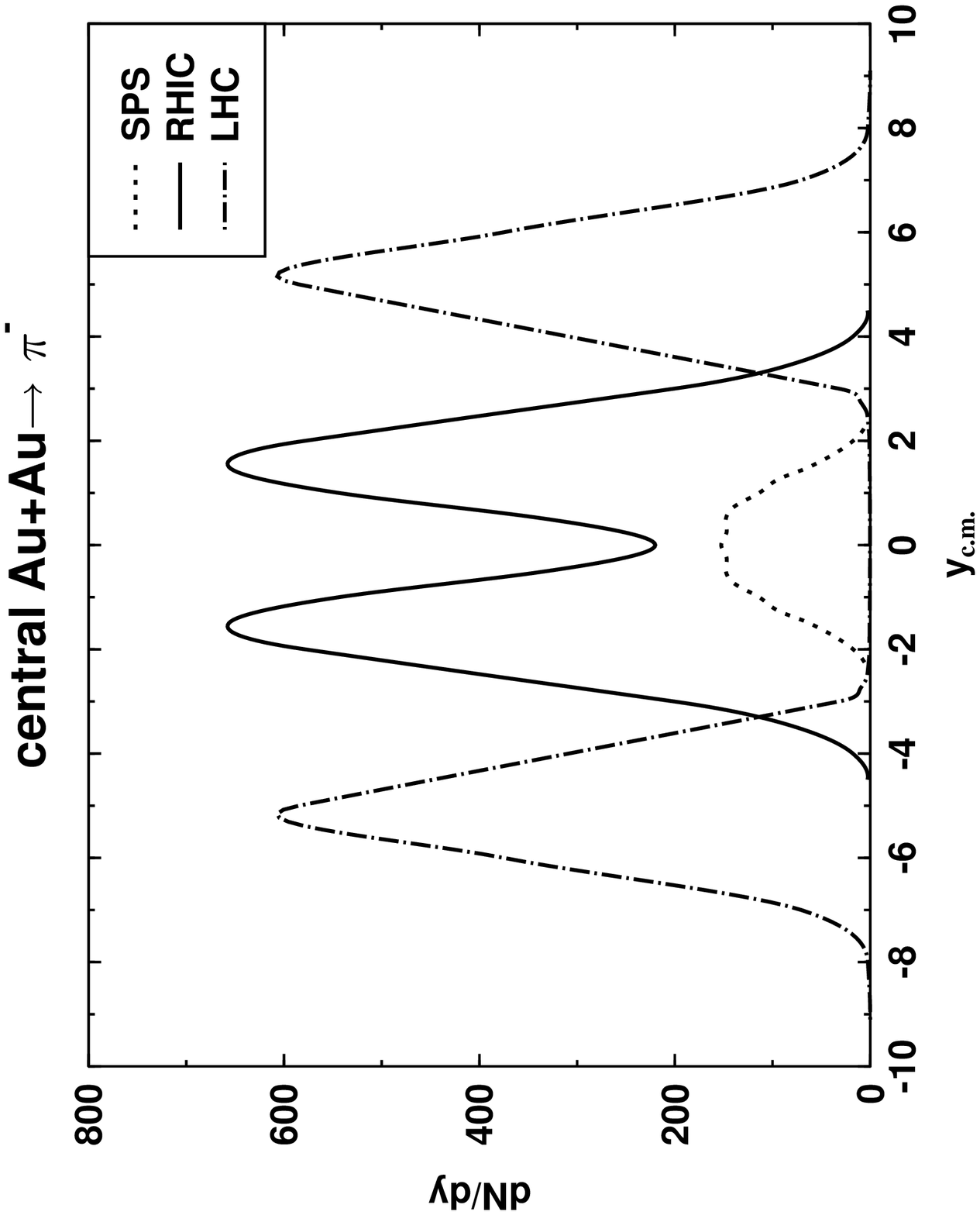} {Figure~3: Rapidity spectra of
$\pi^-$ mesons in central Au+Au collisions at different bombarding energies.}
\hspace{2cm}
\putf{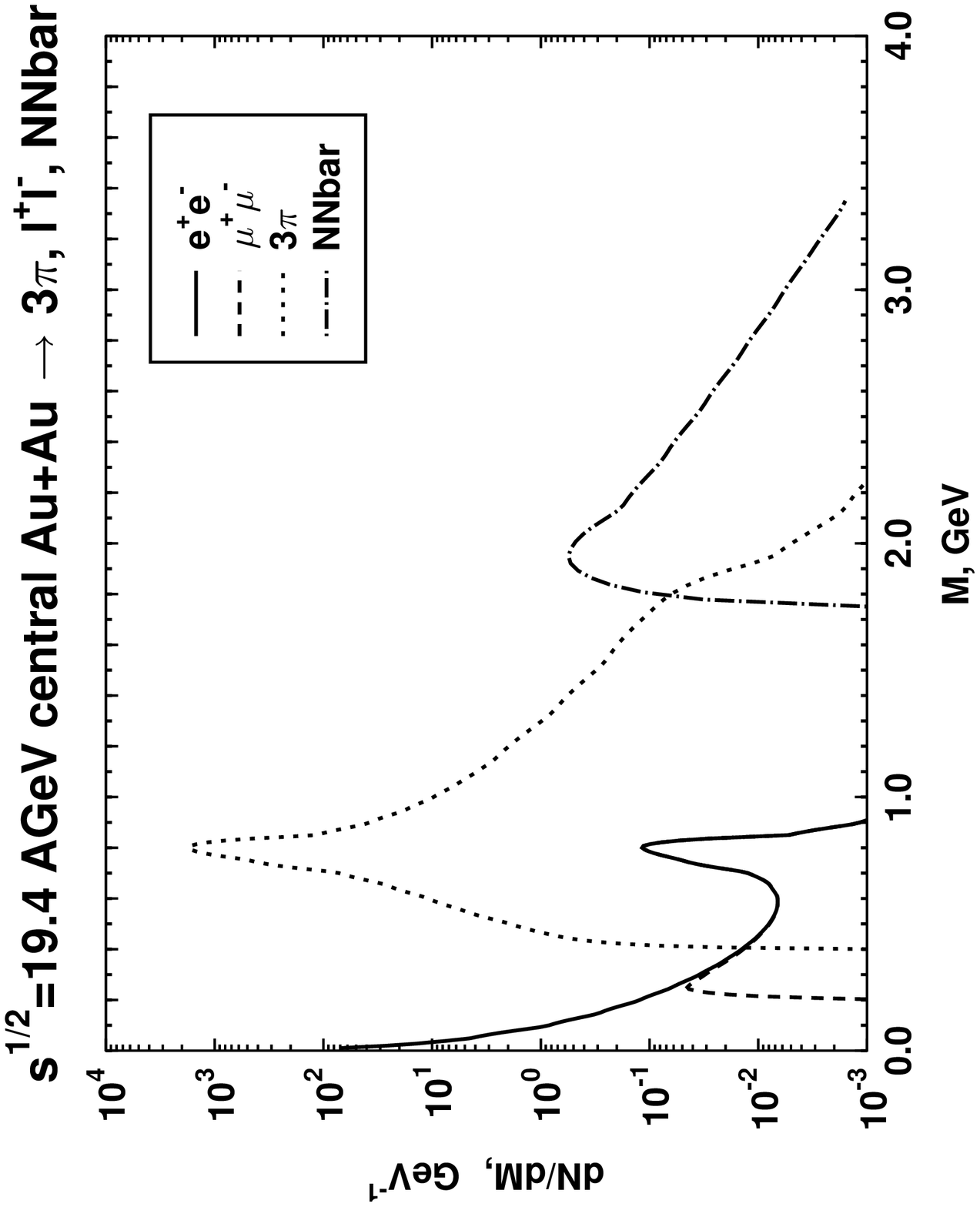}{\mbox
{Figure~4: Distributions over} inva\-riant masses of particles
in different decay channels of virtual $\omega$
mesons produced in central Au+Au collision at SPS energy.}
\\

Fig.~4 shows the distributions over invariant masses of particles produced
in different channels of bremsstrahlung conversion in the case
of a central Au+Au collision at the SPS energy. Note that the mass spectrum
of e$^+$e$^-$ pairs created by the bremsstrahlung mechanism has a strong peak
at invariant masses below the $\omega$ meson mass. On the
other hand, attempts to explain the low mass dilepton yield by the
conventional incoherent mechanisms (e.g. due to the
$\pi\pi\rightarrow\rho\rightarrow{\rm e}^+{\rm e}^-$ processes) strongly
underestimate the observable data~\cite{Cer95}.\\


\section{Conclusions}
In this work we have shown that the coherent bremsstrahlung of the vector
meson field may be an important source of particle production already
at the SPS bombarding energies. The observable signals of this mechanism
may be the two--hump structure of pion and antibaryon rapidity
spectra as well as the enhanced yield of low mass dileptons.
The sharp energy and A--dependence of pion, dilepton and antibaryon
excitation functions can be also a signature of the considered mechanism.
The latter may be responsible, at least partly, for a rapid increase
of the pion multiplicity observed in transition from the AGS to SPS
energy~\cite{Gaz95}.

\section*{Acknowledgments}

The authors thank Yu.B.~Ivanov, S.~Schramm and L.A.~Winckelmann for valuable
discussions.  This work has been supported in part by the EU--INTAS Grant
No.~94--3405.  We acknowledge also the financial support from GSI, BMFT and
DFG.

\end{document}